\documentclass[usenatbib]{mn2e} 
\usepackage{epsfig}

\newif\ifAMStwofonts
\AMStwofontstrue

\def\be{\begin{equation}}
\def\ee{\end{equation}}

\def\gtsima{$\; \buildrel > \over \sim \;$}
\def\ltsima{$\; \buildrel < \over \sim \;$}
\def\prosima{$\; \buildrel \propto \over \sim \;$}
\def\gsim{\lower.5ex\hbox{\gtsima}}
\def\lsim{\lower.5ex\hbox{\ltsima}}
\def\simgt{\lower.5ex\hbox{\gtsima}}
\def\simlt{\lower.5ex\hbox{\ltsima}}
\def\simpr{\lower.5ex\hbox{\prosima}}

\def\ie{{\frenchspacing\it i.e. }}
\def\eg{{\frenchspacing\it e.g. }}
\title[Ly$\alpha$ heating]{Ly$\alpha$ heating and its impact on early structure formation}

\author[Ciardi \& Salvaterra]{Ciardi B.$^1$ \& Salvaterra R.$^2$\\
$^1$ Max-Planck-Institut f\"ur Astrophysik, Karl-Schwarzschild-Stra\ss e 1, 85748 Garching, Germany\\
$^2$ Dipartimento di Fisica G. Occhialini, Universit\`a degli Studi di Milano Bicocca, Piazza della
Scienza 3, 20126 Milano, Italy\\} 

\date{April 07}       

\pagerange{\pageref{firstpage}--\pageref{lastpage}}
\pubyear{2007}
\begin{document}

\maketitle
\label{firstpage}

\begin{abstract}
In this paper we have calculated the effect of Ly$\alpha$ photons emitted by the first
stars on the evolution of the IGM temperature. We have considered both a
standard Salpeter IMF and a delta-function IMF for very massive stars with mass 300~M$_\odot$.
We find that the Ly$\alpha$ photons produced by the stellar populations considered here are able
to heat the IGM at $z \simlt 25$, although never above $\sim 100$~K.
Stars with a Salpeter IMF are more effective as, due to the contribution from small-mass 
long-living stars, they produce a higher Ly$\alpha$ background.
Ly$\alpha$ heating can affect the subsequent formation
of small mass objects by producing an entropy floor that may limit the amount of gas able
to collapse and reduce the gas clumping. 
We find that the gas fraction in halos of mass below $\sim 5\times 10^6$~M$_\odot$
is less than 50\% (for the smallest masses this fraction drops to 1\% or less)
compared to a case without Ly$\alpha$ heating. Finally, Ly$\alpha$ photons heat the IGM
temperature above the CMB temperature and render the 21~cm line from neutral hydrogen visible
in emission at $z \simlt 15$.
\end{abstract}

\begin{keywords}
galaxies: formation - feedback - cosmology: theory
\end{keywords}

\section{Introduction}

The study of the high redshift intergalactic medium (IGM) has recently attracted increasing
attention as the new generation of low frequency radio telescopes, \eg 
{\tt LOFAR}\footnote{http://www.lofar.org},
{\tt MWA}\footnote{http://web.haystack.mit.edu/arrays/MWA/}
and {\tt PAST/21cmA}, promises to open an unexplored observational window on 
the high-$z$ universe. In fact, these facilities should be able to detect the 21~cm line associated
with the hyperfine transition of the ground state of neutral hydrogen, either in
absorption or in emission against the Cosmic Microwave Background (CMB) radiation 
(see Furlanetto, Oh \& Briggs 2007 for a review on the topic). In order for the 
line to be visible, the spin temperature, which regulates the population of
the levels, needs to be decoupled from the CMB temperature. While at $z \simgt 20$
collisions efficiently decouple the two temperatures, at lower redshift 
scattering with the Ly$\alpha$ photons emitted by the first stars is the
most efficient process (Wouthuysen 1952; Field 1959; Hirata 2006; Pritchard \& Furlanetto 2006). 
In addition, Ly$\alpha$ photons could be able to heat the IGM temperature above
the CMB temperature and render the 21~cm line visible in emission. The
advantage of Ly$\alpha$ photons over other heating sources, as \eg x-rays 
(unless the spectrum is very hard), is that the Ly$\alpha$
photons can travel cosmological distances and quickly build up a more homogeneous
background.

The impact of Ly$\alpha$ photon scattering on the evolution of the
IGM temperature
has been recently revised by Chen \& Miralda-Escud\`e (2004) and Chuzhoy \& Shapiro (2007).
Chen \& Miralda-Escud\`e (2004) included atomic thermal motion 
in their calculations, finding a heating rate several orders of magnitude lower than
the previous estimate by Madau, Meiksin \& Rees (1997). 
In addition, while ``continuum'' photons (with frequency
between the Ly$\alpha$ and Ly$\beta$) heat the gas, ``injected'' photons (which
cascade into the Ly$\alpha$ from higher atomic resonances) cool the gas, resulting
in an effective cooling from Ly$\alpha$ photons at temperatures above 10~K. 
Chuzhoy \& Shapiro (2007) though, showed that the cascade which follows absorption of
photons in resonances higher than the Ly$\alpha$ happens via the 2s level rather than
the 2p level. Thus, the number of ``injected'' photons and their cooling
efficiency are reduced compared to the estimate of Chen \& Miralda-Escud\`e (2004)
and Ly$\alpha$ photon scattering can be an efficient heating
source also at temperatures higher than 10~K.
As a consequence, Ly$\alpha$ photons
could heat the IGM above the CMB temperature and render the
IGM visible in 21~cm line emission. 

The IGM heating from Ly$\alpha$ photons can also have a feedback effect on structure 
formation. In the standard cosmological scenario, the first objects
to collapse have typically small masses, corresponding to virial temperature
$T_{vir} \simlt 10^4$~K. The formation and evolution of these structures is
heavily affected by feedback effects (for a review see Ciardi \& Ferrara 2005).
Chuzhoy \& Shapiro (2007) have noted that objects with virial temperatures of
few thousands degrees forming in an IGM pre-heated by Ly$\alpha$ photons,
would have a reduced post-collapse overdensity and would be more vulnerable to
feedback effects such as molecular hydrogen dissociation and photoevaporation.

All this makes it worthwhile to investigate further the effect of Ly$\alpha$ 
photons on the IGM temperature evolution.
The paper is organized as follows. In Section~2 we estimate the Ly$\alpha$ background
expected from very high redshift stars; in Sections~3 and~4 we describe their effect
on the IGM temperature evolution and feedback on early structure formation; in
Section~5 we discuss the impact on emission/absorption of 21~cm line and in Section~6
we give our Conclusions. Throughout the paper we assume a $\Lambda$CDM cosmology
with $\Omega_\Lambda=0.76$, $\Omega_m=0.24$, $\Omega_b=0.04$, $h=0.73$ and $\sigma_8=0.85$. 

\section{Ly$\alpha$ photon production}

In this Section we estimate the Ly$\alpha$ background, $J_\alpha$, expected from the
first stars.
Dark matter decays and annihilations provide an additional contribution to $J_\alpha$
(e.g. Valdes et al. 2007), but this becomes negligible once
the first sources of radiation form (Valdes, private communication). For this reason 
we limit our estimate to the stellar component.
At each redshift, the comoving specific emissivity from such stars, $\epsilon_\nu$, is given by: 

\begin{equation}
\epsilon_\nu(z)=\int_z^{\infty} dz^\prime l_\nu(t_{z,z^\prime}) 
\int^{\infty}_{M_{min}(z^\prime)} M_\star \frac{d^2 n}{dM_h dz^\prime}(M_h,z^\prime) dM_h,
\label{eq:ep}
\end{equation}

\noindent
where $d^2n/dM_h dz$ is the formation rate of halos of total 
mass $M_h$ as given by the Press-Schechter formalism (Press \& Schechter 1974),
$M_\star$ is the corresponding stellar mass and $M_{min}$ is the minimum 
mass of the halos. In this work we assume that 
stars can form in H$_2$ cooling halos with virial temperature as low as $400$ 
K and for $M_{min}$ we use the parameterization adopted by Santos, Bromm \& 
Kamionkowski (2002), \ie $M_{min}(z)\sim 2.6\times 10^5\; {\rm M}_\odot \;[(1+z)/10]^{-3/2}$. 
The stellar mass is related to the total halo mass by: 

\begin{equation}
M_\star=f_\star f_{gas} \frac{\Omega_b}{\Omega_m} M_h,
\label{eq:mstar}
\end{equation}

\noindent
where $f_\star$ is the star formation efficiency and 
$f_{gas}$ is the fraction of baryons able to collapse
(see Sec.~4). In our fiducial case $f_\star=0.1$ and $f_{gas}=1$.
$l_\nu(t_{z,z^\prime})$ is the template specific luminosity for a
stellar population of age $t_{z,z^\prime}$ (time elapsed between redshift 
$z^\prime$ and $z$). We adopt the stellar spectra for metal-free stars 
computed by Schaerer (2002) and two different Initial Mass Functions (IMFs):
a Salpeter IMF with lower/upper cut-off mass of 1/100~M$_\odot$,
and a delta-function IMF for Very Massive Stars (VMS) with mass 300~M$_\odot$.
For completeness, we have also considered a case with Pop~II stars and a
Salpeter IMF. Although the ionizing photon production is much reduced
(by a factor of $\sim 4$) compared to metal-free
stars with the same IMF, the Ly$\alpha$ photon emission is similar. For this reason we will
limit our discussion to metal-free stars.

The background intensity $J_{\nu_0}$ seen at a frequency $\nu_0$ by an
observer at redshift $z_0$ is then given by:
\begin{equation}
J_{\nu_{0}} (z_0)= \frac{(1+z_0)^3}{4\pi}\int^{\infty}_{z_{0}}
\epsilon_\nu(z)e^{-\tau_{eff}(\nu_{0},z_{0},z)}\frac{dl}{dz}dz,
\end{equation}

\noindent
where $\nu=\nu_0(1+z)/(1+z_0)$, $dl/dz$ is the proper line element, $\tau_{eff}(\nu_{0},z_{0},z)$ 
is the effective optical depth at $\nu_0$ of the IGM between 
redshift $z_0$ and $z$ (see Sec~2.2 of Salvaterra \& Ferrara 2003 for a full
description of the IGM modeling) and $\epsilon_\nu(z)$ is provided by equation \ref{eq:ep}.

\begin{figure}
\psfig{figure=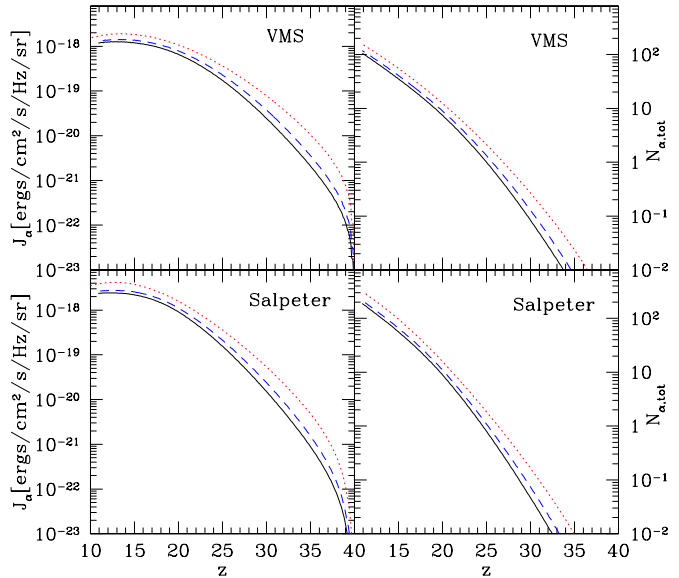,width=0.50\textwidth}
\caption{\label{emiss}Ly$\alpha$ photon background (left panels) and total number of 
photons between the Ly$\alpha$ and the Ly-limit per baryon (right) as a function of 
redshift for VMS (upper panels) and 
Salpeter (lower panels) IMF. In both panels the dotted line refers to the case
without coupling with the feedback effect, while the solid (dashed) line includes
the coupling and a $\delta_{vir} \sim 180$ ($\delta_{vir}=50$)
(see text for details).}
\end{figure}

In Figure~\ref{emiss} we plot (dotted lines) the evolution of the background intensity at the Ly$\alpha$ 
frequency, $J_\alpha$, (left panels) and the total number of photons between the
Ly$\alpha$ and the Ly-limit per baryon emitted by redshift $z$, $N_{\alpha,tot}$, (right).
A case with a VMS and Salpeter IMF is shown in the upper and lower panels respectively.
As expected, although at the highest redshifts a VMS IMF produces more
photons, a Salpeter IMF gives a higher photon density at $z \simlt 26$ 
because of the contribution from the small-mass long-living stars.

\section{IGM temperature evolution}

In this Section we calculate the evolution of the IGM temperature, $T_{IGM}$, induced
by Ly$\alpha$ photon heating. 
Note that our calculations assume that the gas is in
a neutral state, \ie the filling factor of ionized gas due to reionization is small.
Just for reference, assuming an escape fraction of ionizing photons of 10\%, we find
that 10 photons per baryon are emitted in the redshift range (depending on other parameters
adopted) 12-14 (18-20) for a Salpeter (VMS) IMF and 8-10 for Pop~II stars with a Salpeter IMF.

$T_{IGM}$ is calculated with {\tt RECFAST} (Seager, Sasselov \& Scott 1999; 2000) until the 
first sources of radiation
turn on. Subsequently, we follow the method illustrated by Chuzhoy \& Shapiro (2007). Here,
we briefly summarize the relevant equations and we refer the reader to the original paper
for more details. As the authors show that the presence of deuterium can affect the IGM
temperature (its influence increasing with decreasing redshift and increasing $N_{\alpha,tot}$),
we have included D in our calculations. 
When Ly$\alpha$ resonance photons are the major heating source, the H and
D temperature evolve as:
\begin{equation}
\frac{dT_{\rm H}}{dt}=\frac{2}{3k} \left(H_\alpha+L_{\rm D} \frac{n_{\rm D}}{n_{\rm H}} \right)
- \frac{4T_{\rm H}}{3t},
\end{equation}
\begin{equation}
\frac{dT_{\rm D}}{dt}=\frac{2}{3k} \left(H_\beta+L_{\rm D}\right)
- \frac{4T_{\rm D}}{3t},
\end{equation}
where $L_{\rm D}$ is the D energy loss rate from collisions with H, $n_{\rm H}$ ($n_{\rm D}$) is
the number density of H (D) and $H_\alpha$ ($H_\beta$) is the heating rate per H (D) atom.
The expressions for $L_{\rm D}$, $H_\alpha$ and $H_\beta$ can be found in the original paper 
and a fit for the energy gain/loss from ``continuum''/``injected'' photons has been provided
by L. Chuzhoy (Chuzhoy private communication). 
The rates $\dot{N}_\alpha$ and $\dot{N}_{\beta}$ (the number of photons that pass through the
Ly$\alpha$ and Ly$\beta$ resonance per H atom per unit time, respectively)
necessary to calculate $H_\alpha$ and $H_\beta$ have been derived from the 
model described in the 
previous Section. Our fiducial calculations assume a ratio of ``injected'' to ``continuum''  
photons $J_i/J_c=0.15$ and a deuterium abundance of [D/H]=$5 \times 10^{-5}$.

\begin{figure}
\psfig{figure=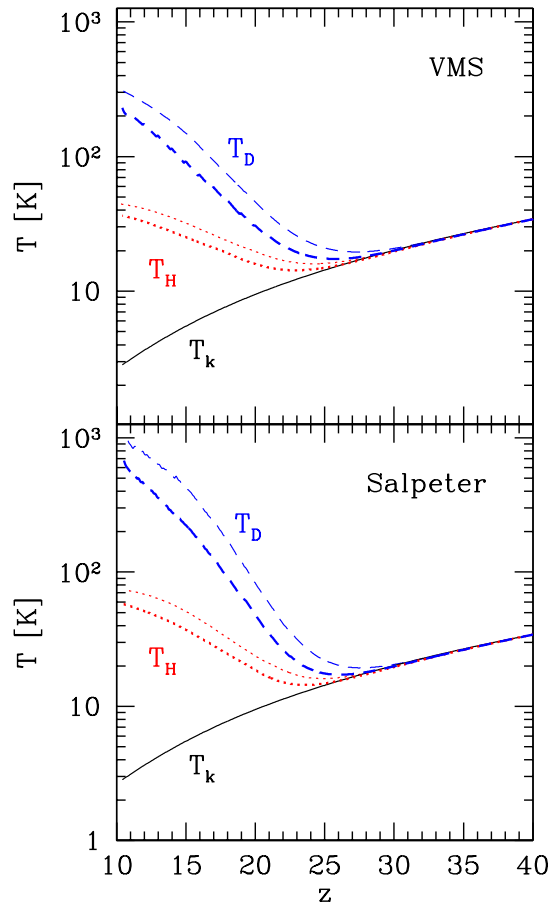,width=0.50\textwidth}
\caption{\label{temp} Temperature evolution for VMS (upper panel) and Salpeter (lower 
panel) IMF. The IGM temperature in the absence of Ly$\alpha$ heating, $T_k$, is plotted
(solid lines) together with $T_{\rm H}$ (dotted) and $T_{\rm D}$ (dashed). Thick (thin)
lines refer to the case in which the feedback effect from Ly$\alpha$ photons is included
(excluded) (see text for details).}
\end{figure}

In Figure~\ref{temp} the temperature evolution is plotted as a function of redshift for a
VMS (upper panel) and Salpeter (lower) IMF. The solid line represents the IGM temperature
calculated with {\tt RECFAST} in the absence of Ly$\alpha$ heating, $T_k$. $T_{\rm H}$ 
($T_{\rm D}$) is plotted with thin dotted (dashed) lines. As expected, the temperatures are
higher with a Salpeter IMF, reflecting the behavior
of the Ly$\alpha$ photon production (see Fig.~\ref{emiss}). As discussed in Chuzhoy \&
Shapiro (2007), the most relevant resonance for deuterium is Ly$\beta$. They show that
D Ly$\beta$ photons have a negative color temperature and as a consequence,
they are much more efficient than Ly$\alpha$ photons in heating the gas. In addition,
differently from H Ly$\alpha$ photons, D Ly$\beta$ photons efficiency as heaters increases
with temperature. As a result, $T_{\rm D}$ is higher than $T_{\rm H}$. We also
find that the
value of $T_{\rm H}$ is independent of the presence of D. In the following, we will refer
to $T_{\rm H}$ indifferently as the hydrogen temperature or the IGM temperature, $T_{\rm IGM}$.
 
To study the effect of the parameters adopted in our fiducial run, we
have done additional calculations varying: {\it (i)} the abundance of D in the range [D/H]=
$5 \times 10^{-6}-5 \times 10^{-4}$; {\it (ii)} the ratio of ``injected'' to ``continuum''
photons in the range $J_i/J_c=0.1-0.2$; {\it (iii)} the star formation efficiency $f_\star=0.05-0.2$.
We find that although $T_{\rm D}$ 
strongly depends on [D/H] ($H_\beta$ increases 
if [D/H] decreases), the presence of D does not affect $T_{\rm H}$.
On the other hand, $T_{\rm H}$ slightly increases with decreasing $J_i/J_c$ (because
the cooling from the ``injected'' photons is less effective),
but the changes are too small to be relevant for this study.
A lower (higher) value of the star formation efficiency induces a smaller 
(larger) $J_\alpha$ and gas temperature. 

\section{Feedback from Ly$\alpha$ photons}

In this Section we will discuss the feedback effect of Ly$\alpha$ heating on the formation
of small mass objects.

To study the impact of Ly$\alpha$ feedback we follow the method outlined by Oh \& Haiman (2003)
and compare the IGM entropy, $K_{IGM}=T_{IGM}/n_{IGM}^{2/3}$, with the entropy generated
by gravitational shock heating, $K_0=T_{vir}/n(r_{vir})^{2/3}$, where $n_{IGM}$
is IGM number density, while $T_{vir}$ and $n(r_{vir})$ are the
virial temperature and the number density at the virial radius of a halo.  The authors show
that as the entropy parameter $\tilde{K}=K_{IGM}/K_0$ increases, the Jeans smoothing effects
due to the IGM entropy become more significant, the central pressure and
density of the halos decrease, together with the accreted gas fraction and gas clumping
(see also Gnedin 2000). 
As a consequence, the amount of gas going into stars is likely to decrease and 
the halo becomes more vulnerable to
feedback effects such as \eg molecular hydrogen dissociation and photoevaporation. 
It should be noted that $\tilde{K}$ has a minimum at $r_{vir}$, because of the higher
halo density at $r<r_{vir}$. Throughout this paper we will always be conservative
and refer to the value at the virial radius.

The value of $K_0$ depends on the halo profile and the entropy parameter can be written as:
\begin{equation}
\tilde{K}=\frac{T_{IGM}}{T_{vir}} \delta_{vir}^{2/3},
\end{equation}
where $\delta_{vir}$ is the overdensity at the virial radius. For our fiducial runs we have
adopted the value corresponding to a spherical top-hat collapse (Bryan \& Norman 1998):
\begin{equation}
\delta_{vir}=18 \pi^2+82 (\Omega_m^z -1)-39 (\Omega_m^z -1)^2,
\end{equation}
where:
\begin{equation}
\Omega_m^z=\frac{\Omega_m (1+z)^3}{\Omega_m (1+z)^3+\Omega_\Lambda}.
\end{equation}   
For the cosmology adopted here $\delta_{vir} \sim 180$. 

\begin{figure}
\psfig{figure=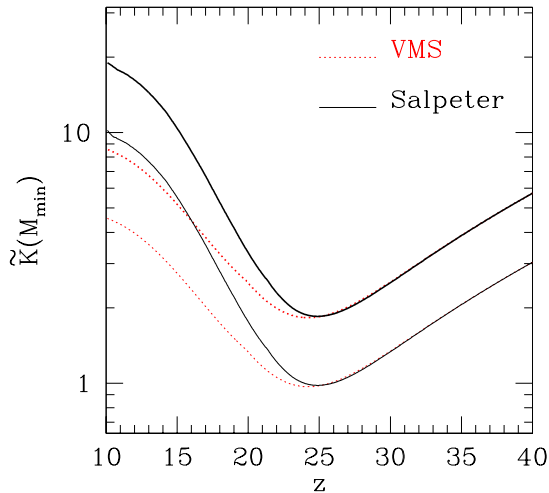,width=0.50\textwidth}
\caption{\label{entropy} Evolution of the entropy parameter corresponding to $M_{min}$ 
for VMS (dotted lines) and Salpeter (solid) IMF. Thick (thin) lines correspond to a
case with $\delta_{vir} \sim 180$ ($\delta_{vir}=50$).}
\end{figure}

To quantify the strength of such feedback, we plot in Figure~\ref{entropy}, as an example,
the redshift evolution of the entropy parameter for halos with mass $M_{min}$.
The thick solid (dotted) lines refer to a Salpeter (VMS) IMF. 
According to Oh \& Haiman (2003) calculations (see \eg their Fig.~6) the fraction of
gas, $f_{gas}$, within the virial radius of a halo
with $\tilde{K}$ larger than 1 is smaller than 50\% and as low as 1\% for $\tilde{K}\sim 10$.
Thus, the fraction of gas collapsing in halos with mass equal to $M_{min}$ can be
effectively reduced to few percents. So, to account for the effect
of Ly$\alpha$ feedback in a self-consistent way, we repeat the
same calculations adopting for $f_{gas}$ in equation~\ref{eq:mstar} the value
corresponding to $\tilde{K}(T_{IGM},M_h)$. This has the effect of lowering the amount
of gas that goes into stars and thus the intensity of the Ly$\alpha$ background.
The evolution of $J_\alpha$ and $N_{\alpha,tot}$ when the coupling between
the Ly$\alpha$ feedback and structure formation is included, is shown in Figure~\ref{emiss}
as solid lines for $\delta_{vir} \sim 180$. As expected,
both $J_\alpha$ and $N_{\alpha,tot}$ are reduced by the effect of feedback.
As a consequence, also the IGM temperature is lower when the coupling is included
(thick lines in Fig.~\ref{temp}).

\begin{figure}
\psfig{figure=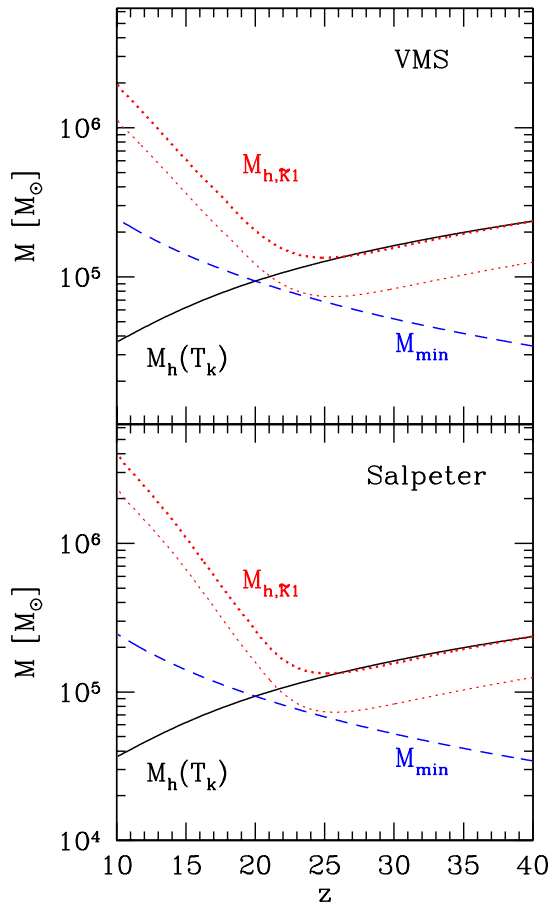,width=0.50\textwidth}
\caption{\label{feed} Mass evolution for VMS (upper panel) and Salpeter (lower
panel) IMF. The halo mass corresponding to $T_k$ is plotted (solid lines) together with
$M_{h, \tilde{K}1}$ (dotted) and $M_{min}$ (dashed). The thick (thin) lines correspond to
$\delta_{vir}\sim 180$ ($\delta_{vir}=50$)(see text for details).}
\end{figure}

Given the IGM temperature, we can 
derive the virial temperature corresponding to $\tilde{K}=1$, $T_{vir, \tilde{K}1}$. 
Halos with $T_{vir}<T_{vir, \tilde{K}1}$ will be heavily affected by Ly$\alpha$ photon feedback. 
From $T_{vir, \tilde{K}1}$
one can derive the corresponding halo mass, $M_{h, \tilde{K}1}$. In Figure~\ref{feed} the mass
evolution is shown as a function of redshift for a VMS (upper panel) and Salpeter (lower) IMF. 
The solid lines represent the halo mass corresponding to $T_k$. $M_{h, \tilde{K}1}$ (thick dotted
lines) is plotted together with $M_{min}$ (dashed). The feedback from Ly$\alpha$ 
photons will greatly affect the formation and evolution of halos with 
mass below $M_{h, \tilde{K}1}$, in which only about 50\% of the gas fraction will be able to 
collapse. The effect is stronger for a Salpeter IMF. It should be noted that 
the Ly$\alpha$ feedback will only affect the collapse of mini-halos, while the formation of
halos collapsing via H-line cooling will proceed unaffected. 

As mentioned above, the value of $\delta_{vir}$ depends on the density profile of the
halo. If a Navarro, Frenk \& White (1997) profile is adopted, the density contrast at the
virial radius for a halo with mass $M_{min}$ varies in the range $\sim 50-70$, depending
on the redshift. As a comparison, we have repeated the same calculations for $\delta_{vir}=50$ and we show
the evolution of  $J_\alpha$ and $N_\alpha$ in Figure~\ref{emiss} 
with dashed lines.
The effect on $\tilde{K}$ and $M_{h, \tilde{K}1}$ can be seen as thin lines in Figures~\ref{entropy} 
and~\ref{feed}.  As expected, the impact of the Ly$\alpha$ feedback is 
reduced compared to the fiducial case, but it can still be appreciable. 

\section{Effect on 21~cm line}

As already discussed in the Introduction, heating of the IGM from Ly$\alpha$ photons has
important consequences for the observation of 21~cm emission line from neutral hydrogen.
As the physics behind the emission/absorption of such line has been discussed extensively
by several authors (for a review see \eg Furlanetto, Oh \& Briggs 2007), here we just write the relevant
equations. 
The evolution of the spin temperature, $T_s$, can be written as (Chuzhoy \& Shapiro 2006):
\begin{equation}
T_s=\frac{T_{\rm CMB} + (y_{\alpha,eff} + y_c) T_k}{1+y_{\alpha,eff}+y_c},
\end{equation}
where $T_{\rm CMB}$ is the CMB temperature;
\begin{equation}
y_c=\frac{T_\star}{A_{10}T_k}(C_{\rm H} + C_e +C_p),
\end{equation}
is the coupling efficiency due to collisions with H atoms, electrons and protons,
$T_\star=0.068$~K is the temperature corresponding to the energy difference between the
singlet and triplet hyperfine levels of the ground state of neutral hydrogen and
$A_{10}=2.85 \times 10^{-15}$~s$^{-1}$ is the spontaneous emission rate. For
the de-excitation rates $C_{\rm H}$, $C_e$ and $C_p$ we have adopted the fits used
by Kuhlen, Madau \& Montgomery (2006; see also the original papers by Smith 1966; Allison
\& Dalgarno 1969; Liszt 2001; Zygelman 2005), \ie $C_{\rm H}= n_{\rm H} \kappa$,
$C_e=n_e \gamma_e$ and $C_p=3.2 n_p \kappa$ with $n_{\rm H}$, $n_e$, $n_p$ hydrogen,
electron and proton number density, $\kappa=3.1 \times 10^{-11}
T_k^{0.357} {\rm exp}(-32/T_k)$~cm$^3$~s$^{-1}$ and ${\rm log}(\gamma_e/1 {\rm cm}^3{\rm s}^{-1})=
-9.607+0.5{\rm log}(T_k) {\rm exp}[-({\rm log}T_k)^{4.5}/1800]$. Finally,
\begin{equation}
y_{\alpha,eff}=y_\alpha {\rm e}^{-0.3(1+z)^{1/2} T_k^{-2/3}} \left( 1+\frac{0.4}{T_k}\right)^{-1},
\end{equation}
is the effective coupling efficiency due to Ly$\alpha$ scattering which
takes into account the back-reaction of the resonance on the Ly$\alpha$ spectrum and the effect of
resonant photons other than Ly$\alpha$, and $y_\alpha=P_{10}T_\star/(A_{10} T_k)$,
with $P_{10}\sim 10^9 J_\alpha$~s$^{-1}$ radiative de-excitation rate due to Ly$\alpha$ photons.

\begin{figure}
\psfig{figure=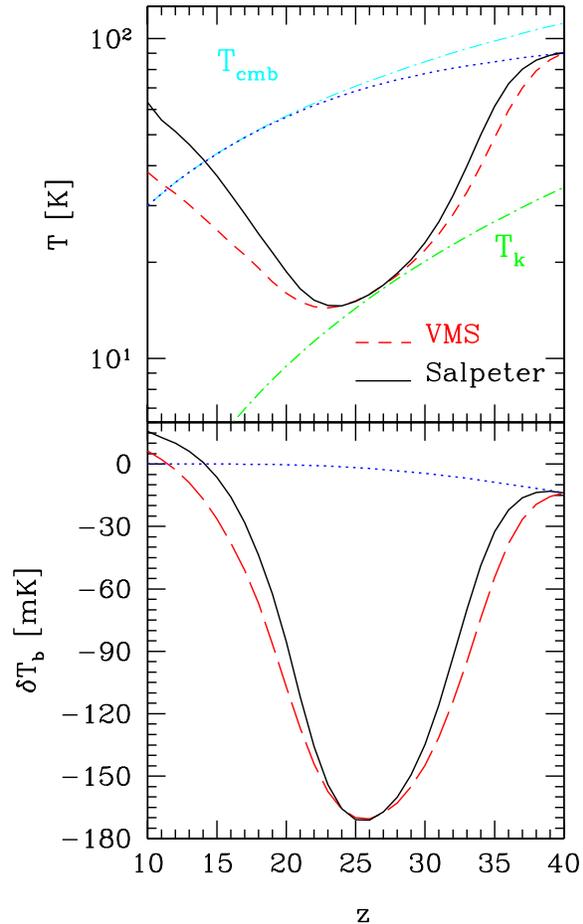,width=0.50\textwidth}
\caption{\label{temp21cm} {\it Upper panel}: Evolution of the spin temperature for a Salpeter (solid
lines) and a VMS (dashed) IMF, and in the absence of a Ly$\alpha$ background (dotted). 
$T_{\rm CMB}$ and $T_k$ are plotted
as upper and lower dotted-dashed lines, respectively. {\it Lower panel}: Evolution of the
differential brightness temperature. Lines are the same as in the upper panel.}
\end{figure}

In the upper panel of Figure~\ref{temp21cm} the spin temperature is shown as a function of redshift
together with $T_{\rm CMB}$ and $T_k$ (upper and lower dotted-dashed lines, respectively).
The dotted line is $T_s$ in the absence of a Ly$\alpha$ background, while the solid (dashed) lines
are for a Salpeter (VMS) IMF with $\delta_{vir} \sim 180$. 
The value of $T_{IGM}$ is hardly affected
by the choice of $\delta_{vir}$, as the difference in the Ly$\alpha$ background is too
small to result in an appreciable difference in temperature. For this reason, in
the following we will discuss only the case with $\delta_{vir} \sim 180$.
In all cases the coupling with the Ly$\alpha$ feedback is included.
In the absence of a Ly$\alpha$ background, $T_s$ gets coupled to $T_{\rm CMB}$ at $z \simlt 20$,
once collisions becomes ineffective in maintaining $T_s=T_k$.
As the intensity required for a Ly$\alpha$ background to be effective in decoupling the spin
temperature from $T_{\rm CMB}$ is 
$J_\alpha \simgt 10^{-22} (1+z)$~ergs~cm$^{-2}$~s$^{-1}$~Hz$^{-1}$~sr$^{-1}$
(\eg Ciardi \& Madau 2003), $T_s$ gets coupled to $T_k$ as early as $z \sim 27$.

The differential brightness temperature, $\delta T_b$, between the CMB and a patch of neutral hydrogen
with optical depth $\tau$ and spin temperature $T_s$ at redshift 
$z$, can be written as:
\begin{equation}
\delta T_b=(1-{\rm e}^{-\tau}) \frac{T_s-T_{\rm CMB}}{1+z}.
\end{equation}
For a gas at the mean IGM density, the optical depth is:
\begin{equation}
\tau(z)=\frac{3A_{10}}{32 \pi \lambda^3} \frac{T_\star}{T_s} \frac{n_{\rm HI}(z)}{H(z)},
\end{equation}
where $\lambda$ is the wavelength of the transition, $n_{\rm HI}$ is the number density 
of neutral hydrogen and $H(z)$ is the Hubble constant.

In the lower panel of Figure~\ref{temp21cm} $\delta T_b$ is plotted for the spin temperature 
calculated above, for a Salpeter (solid lines) and a VMS (dashed) IMF, and in the absence
of a Ly$\alpha$ background (dotted). 
From Figure~\ref{temp21cm} it is clear that, although Ly$\alpha$ photons are very efficient in 
decoupling the spin temperature
from the CMB temperature, they are not able to heat the IGM above $T_{CMB}$ before $z \sim 15$.
Thus, if Ly$\alpha$ photons are the only heating source in the early universe and the 
reionization process is ongoing at very high redshift, we expect to
observe the 21~cm line in emission only when the IGM is substantially ionized, while at earlier
times we expect to observe it in absorption. In the absence of feedback, the heating by
Ly$\alpha$ photons is effective in raising the IGM temperature above $T_{CMB}$ already at
$z \sim 17$.

\section{Discussion and Summary}

In this paper we have calculated the effect of Ly$\alpha$ photons emitted by the first 
metal-free stars on the evolution of the IGM temperature. We have considered both a
standard Salpeter IMF and a delta-function IMF for very massive stars with mass 300~M$_\odot$.
Our calculations are run for a neutral gas at the mean density, thus they are strictly valid
as long as the filling factor of ionized gas is small or in regions of the IGM that
have not been reached by ionizing radiation. 

We find that the Ly$\alpha$ photons produced by the stellar populations considered here are able 
to heat the IGM at $z \simlt 25$, although never above $\sim 100$~K (thin lines in Fig.~\ref{temp}). 
Stars with a Salpeter IMF are more effective as, due to the contribution from small-mass long-living stars,
they produce a higher Ly$\alpha$ background.

Ly$\alpha$ heating can affect the subsequent formation
of small mass objects by producing an entropy floor that may limit the amount of gas able
to collapse. As shown in Oh \& Haiman (2003), as the entropy parameter increases, the Jeans
smoothing effects due to the IGM heating become more significant, the central pressure and
density of halos decreases as well as the accreted gas fraction and clumping, with the consequence
of rendering these halos more vulnerable to \eg molecular hydrogen dissociation and photoheating.
We find that the gas fraction in halos of mass in the range $10^5-5 \times 10^6$~M$_\odot$ 
(depending on redshift) is less than 50\% (for the smallest masses this fraction drops to 1\% or less)
compared to a case without Ly$\alpha$ heating. It should be noted that 
H-cooling halos (i.e. $T_{vir}\ge 10^4$ K) are not affected by this 
feedback.

%only halos with masses
%$\simlt 10^8$~M$_\odot$ are affected by this feedback.

To include the effect of feedback in a self-consistent way, we have repeated the same calculations
adopting for the fraction of gas able to collapse in a halo, instead of unity, the value induced
by the feedback effect. This results in an overall lower amount of collapsed gas and 
a lower value of the Ly$\alpha$ background and IGM temperature (thick lines in Fig.~\ref{temp}).
To have a better understanding of the effects of Ly$\alpha$ heating on structure formation one should
follow the collapse of a halo and determine the impact on the gas density profile and clumping, but
this is beyond the scope of this study.

Many other feedback effects affect the formation and evolution of small mass halos in the early
universe (see Ciardi \& Ferrara 2005 for a review), ranging from photoionization/evaporation, to
molecular hydrogen dissociation, to blowout/blowaway. Given the wide variety of feedback effects
and the fact that often they are studied in specific cases, it is difficult to assess the
relative importance of Ly$\alpha$ feedback. Differently from other feedback effects
though (with the exception of molecular hydrogen dissociation), Ly$\alpha$ photons can travel
cosmological distances and quickly build up a fairly homogeneous background. 

Minihalo suppression from Ly$\alpha$ feedback may also have an important 
impact  on the reionization history. The lower star formation rate due to 
reduced collapse gas mass results in a delay of the process. On the other hand, Ly$\alpha$ 
feedback can have the opposite effect. The halo profiles are smoothed out and the gas
clumping factor is consequently reduced, so that the reionization process may 
be speeded up by a faster photoevaporation of minihalos. The final effect on the reionization 
history requires dedicated studies (see \eg Shapiro, Iliev \& Raga 2004, 
Ciardi et al. 2006) and is beyond the scope of our analysis.

Once the IGM is heated by the Ly$\alpha$ photons
above the CMB temperature, the 21~cm line associated with the hyperfine transition of the ground
state of neutral hydrogen can be visible in emission against the CMB. This happens
at $z \simlt 15$, while at higher redshift, in the absence of other heating sources as \eg
x-rays, we expect to observe the line in absorption.
Thus, for the detection of the emission line, the reionization history becomes critical as the
peak of the emission (in terms of fluctuations of the differential brightness
temperature) is expected when roughly half of the volume is ionized 
but the ionized regions are not completely overlapped (see \eg Ciardi \& Madau 2003). 
To investigate this issue in more detail the reionization process should be studied 
together with the effect of Ly$\alpha$ photons.
It should be noted that metal-enriched stars would induce the same effect on the IGM temperature
as metal-free stars with a Salpeter IMF. In this case though, the reionization process is
delayed due to the much lower ionizing photon budget and the 21~cm line can be seen in emission over a
longer redshift range. The same comment applies when a transition from an initially 
metal-free stellar population to metal-enriched stars takes place (see \eg Schneider et al. 2006).

To summarize the main results of this work, we find that:
\begin{itemize}
\item the Ly$\alpha$ background from the first stars can heat the IGM gas up to
temperatures of $\sim 100$~K, depending on redshift and parameters adopted;
\item the associate feedback lowers the gas fraction collapsed in halos of mass below $\sim 5\times 10^6$~M$_\odot$
to less than 50\% (for the smallest halos the fraction drops to 1\% or less) compared to  
the case without Ly$\alpha$ heating;
\item Ly$\alpha$ photons heat the IGM temperature above the CMB temperature and render the
21~cm line from neutral hydrogen visible in emission at $z \simlt 15$. If Ly$\alpha$ feedback on
small structures is neglected emission is already possible at $z \sim 17$.
\end{itemize}

\section*{Acknowledgments}
The authors are grateful to P. Madau for useful discussion, L. Chuzhoy for providing a fit
for the cooling/heating rates and M. Valdes for an estimate of the Ly$\alpha$ background
from dark matter decays/annihilations. We also thank an anonymous referee for helpful comments.

\label{lastpage}

\end{document}